\documentclass{eas}
\usepackage{graphicx}
\begin{document}

\TitreGlobal{Mass Profiles and Shapes of Cosmological Structures}

\title{Weak lensing constraints on galaxy halos}
\author{Henk Hoekstra}
\address{Department of Physics \& Astronomy, University of Victoria}
\runningtitle{Weak lensing constraints on galaxy halos}
\setcounter{page}{23}
\index{Hoekstra, H..}
%
\begin{abstract} Weak gravitational lensing has become an important tool 
to study the properties of dark matter halos around galaxies, thanks
to the advent of large panoramic cameras on 4m class telescopes. This
area of research has been developing rapidly in the past few years,
and in these proceedings we present some results based on the
Red-Sequence Cluster Survey, thus highlighting what can be achieved with
current data sets. We present results on the measurement of virial masses as a
function of luminosity and the extent of dark matter halos. Much
larger surveys are underway or planned, which will result in an
impressive improvement in the accuracy of the measurements. However,
the interpretation of future results will rely more and more on
comparison with numerical simulations, thus providing direct tests of
galaxy formation models.
\end{abstract}
\maketitle

\section{Introduction}
Observations of the dynamics of stars and gas in galaxies have
provided important evidence for the existence of dark matter halos
around galaxies. These studies have also shown that tight relations
exist between the baryonic and dark matter components. The latter
findings provide important constraints for models of galaxy formation,
as their origin needs to be explained.

However, dynamical methods require visible tracers, which typically
can be observed only in the central regions of galaxies, where baryons
are dynamically important. In this regime, the accuracy of simulations
is limited and the physics complicated. Hence the interpretation of
observations is complicated and one needs to proceed cautiously.  In
addition assumptions about the orbit structure need to be made.
Instead, it would be more convenient to have observational constraints
on quantities that are robust (both observationally and theoretically)
and easily extracted from numerical simulations. An obvious quantity
of interest is the virial mass of the galaxy.

Fortunately, in recent years it has become possible to probe the outer
regions of galaxy dark matter halos, either through the dynamics of
satellite galaxies (e.g., Prada et al. 2003) or weak gravitational
lensing. In these proceedings we focus on the latter approach, which
uses the fact that the tidal gravitational field of the dark matter
halo introduces small coherent distortions in the images of distant
background galaxies. This signal can nowadays be easily detected in
data from large imaging surveys. It is important to note, however,
that weak lensing cannot be used to study individual galaxies,
but ensemble averaged properties instead.

Since the first detection of this so-called galaxy-galaxy lensing
signal by Brainerd et al. (1996), the significance of the measurements
has improved dramatically, thanks to new wide field CCD cameras on a
number of mostly 4m class telescopes. This has allowed various groups
to image large areas of the sky, yielding the large numbers of lenses
and sources needed to measure the lensing signal. Results from the
Sloan Digital Sky Survey (SDSS) provided a major improvement (e.g.,
Fisher et al. 2000; McKay et al. 2001) over early studies.  Apart from
the increased surveyed area, an important advantage of the more recent
SDSS studies (McKay et al. 2001; Guzik \& Seljak 2002) is the
availability of (photometric) redshift information for the lenses and
sources. This has enabled studies of the dark matter properties as a
function of baryonic content.

Here, we highlight recent progress by presenting results from the
Red-Sequence Cluster Survey (RCS; Gladders \& Yee 2005). Recently
Hsieh et al. (2005) derived photometric redshifts for a subset of the
RCS and we use these results to study the virial mass as a function of
luminosity. We also present measurements of the extent of dark matter
halos and discuss measurements of their shapes. We conclude by
discussing what to expect in the near future, when much larger
surveys start producing results.

\begin{figure}[ht]
\centering
\includegraphics[width=12cm]{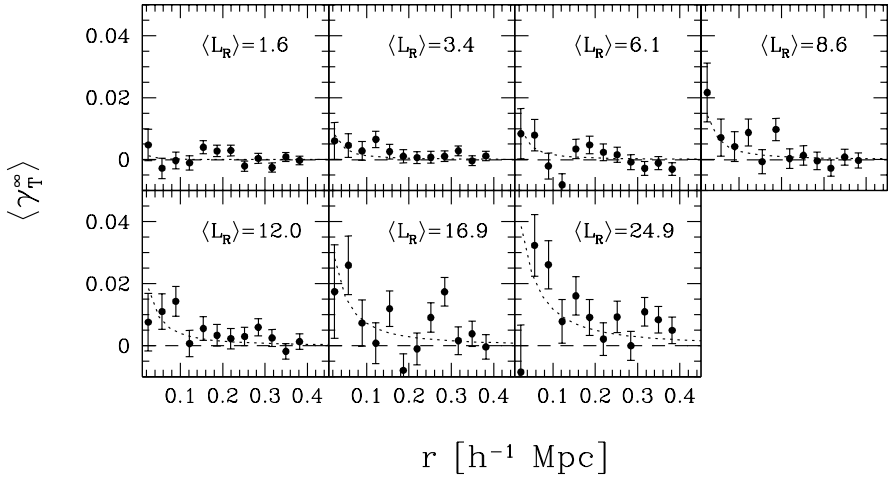}
\caption{\footnotesize Tangential shear as a function of projected
(physical) distance from the lens for each of the seven restframe
$R$-band luminosity bins. To account for the fact that the lenses have
a range in redshifts, the signal is scaled such that it corresponds to
that of a lens at the average lens redshift ($z\sim 0.32$) and a
source redshift of infinity The mean restframe $R$-band luminosity for
each bin is also shown in the figure in units of $10^9
h^{-2}$L$_{R\odot}$. The strength of the lensing signal clearly
increases with increasing luminosity of the lens. The dotted line
indicates the best fit NFW model to the data.
\label{gtprof}}
\end{figure}

\section{Virial masses}

One of the major advantages of weak gravitational lensing over most
dynamical methods is that the lensing signal can be measured out to
large projected distances from the lens. However, at large radii, the
contribution from a particular galaxy may be small compared to its
surroundings: a simple interpretation of the measurements can only be
made for `isolated' galaxies. What one actually observes, is the
galaxy-mass cross-correlation function. This can be compared directly
to models of galaxy formation (e.g., Tasitsiomi et
al. 2004). Alternatively, one can attempt to select only isolated
galaxies or one can deconvolve the cross-correlation function, while
making some simplifying assumptions.  In this section we discuss
results for `isolated' galaxies, whereas in the next section, which
deals with the extent of dark matter halos, we use the deconvolution
method.

A detailed discussion of the results presented in this section can be
found in Hoekstra et al. (2005). The measurements presented here are
based on a subset of the RCS for which photometric redshifts were
determined using $B,V,R_C,z'$ photometry (Hsieh et al. 2005). We
selected galaxies with redshifts $0.2<z<0.4$ and $18<R_C<24$,
resulting in a sample of $\sim 1.4\times 10^5$ lenses. However, to
simplify the interpretation of the results, we proceed by selecting
`isolated' lenses. To do so, we only consider lenses that are at least
30 arcseconds away from a brighter galaxy (see Hoekstra et al., 2005
for details). Note, that bright galaxies are not necessarily isolated.
For such galaxies, however, we expect the lensing signal to be
dominated by the galaxy itself, and not its fainter companions.

\begin{figure}[ht]
\centering
\includegraphics[width=12cm]{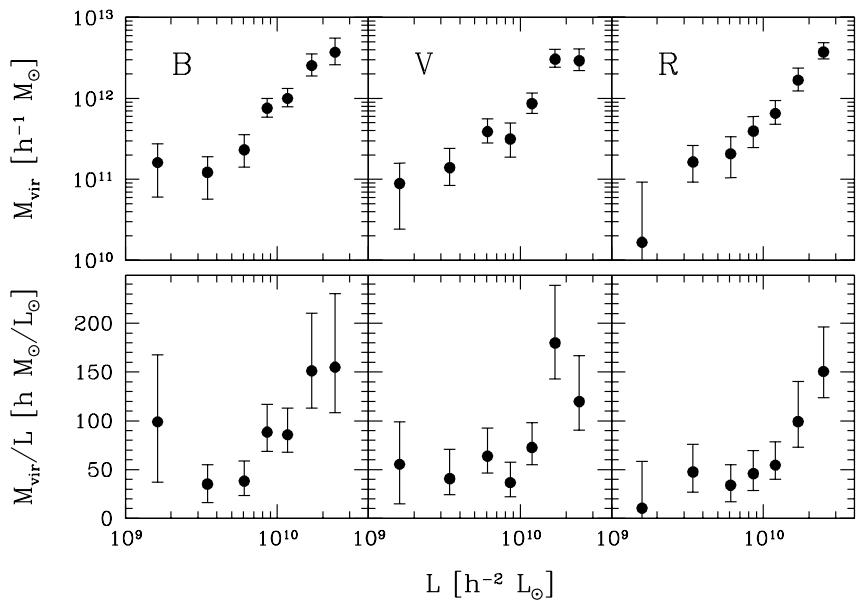}
\caption{\footnotesize {\it upper panels}: Virial mass as a function
of the rest-frame luminosity in the indicated filter. The dashed line
indicates the best fit power-law model for the mass-luminosity
relation, with a power slope of $\sim 1.5$.{\it lower panels:}
Observed rest-frame virial mass-to-light ratios. The results suggest a
rise in the mass-to-light ratio with increasing luminosity, albeit
with low significance.
\label{ml_all}}
\end{figure}

We split the sample into 7 luminosity bins and measure the mean
tangential distortion out to 2 arcminutes from the lens. The resulting
tangential shear profiles are shown in Figure~\ref{gtprof} for the
bins of increasing rest-frame $R$ luminosity. The results for the $B$
and $V$ band are very similar. We estimate the virial mass for each
bin by fitting a a NFW (Navarro, Frenk \& White 1996) profile to the
signal. The resulting virial mass as a function of rest-frame
luminosity is presented in Figure~\ref{ml_all}.  These findings
suggest a power-law relation between the mass and the luminosity,
although this assumption might not hold at the low luminosity end. We
fit a power-law model to the measurements and find that the slope is
$\sim 1.5\pm0.2$ for all three filters.  This results is in good
agreement with results from the SDSS (Guzik \& Seljak, 2002) and
predictions from models of galaxy formation (Yang et al. 2003). As
stressed by Guzik \& Seljak (2002), the observed slope implies that
rotation curves must decline substantially from the optical to the
virial radius.

For a galaxy with a luminosity of $10^{10}h^{-2}L_{B\odot}$ we obtain
a virial mass of $M_{\rm vir}=9.9^{+1.5}_{-1.3}\times 10^{11}h^{-1}M_\odot$.
We note that if the mass-luminosity relation has an intrinsic scatter,
our mass estimates are biased low (Tasitsiomi et al. 2004). The amplitude
of this bias depends on the assumed intrinsic scatter. The results
presented in Tasitsiomi et al. (2004), however, do indicate that the
slope of the mass-luminosity relation is not affected.

\section{Extent and shapes of halos}

The galaxy-mass cross-correlation function is the convolution of the
galaxy distribution and the underlying galaxy dark matter profiles.
Provided we have a model for the latter, we can `predict' the expected
lensing signal. Such an approach naturally accounts for the presence
of neighbouring galaxies. It essentially allows us to deconvolve the
galaxy-mass cross-correlation function, under the assumption that all
clustered matter is associated with the lenses. If the matter in
galaxy groups (or clusters) is associated with the halos of the group
members (i.e., the halos are indistinguishable from the halos of
isolated galaxies) our results should give a fair estimate of the
extent of galaxy halos. However, if a significant fraction of the dark
matter is distributed in common halos, a simple interpretation of the
results becomes more difficult.

\begin{figure}[h]
\centering
\includegraphics[width=10.5cm]{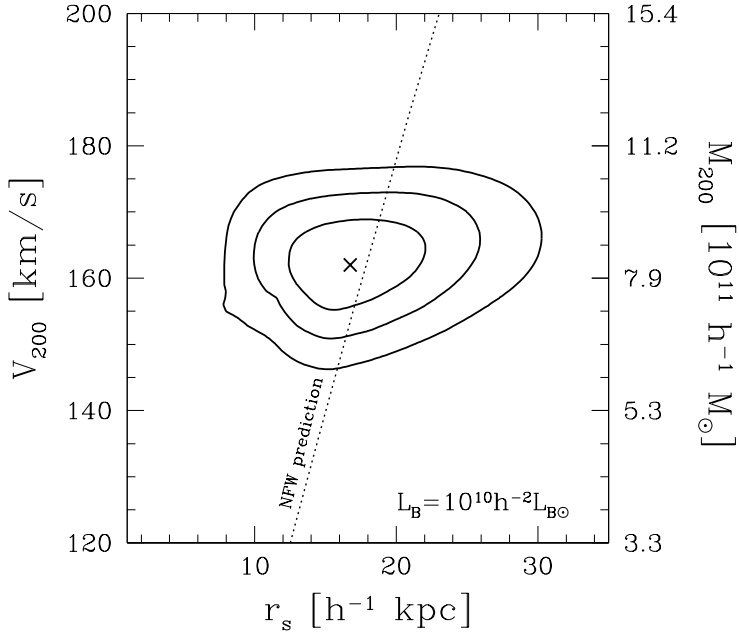}
\caption{\footnotesize Joint constraints on $V_{200}$ and scale radius
$r_s$ for a fiducial galaxy with $L_{\rm B}=10^{10}h^{-2}L_{{\rm
B}\odot}$, with an NFW profile. The corresponding values for $M_{200}$
are indicated on the right axis. The contours indicate the 68.3\%,
95.4\%, and the 99.7\% confidence on two parameters jointly. The cross
indicates the best fit value. The dotted line indicates the
predictions from the numerical simulations, which are in excellent
agreement with our results.
\label{size_nfw}}
\end{figure}

We use such a maximum likelihood approach to place constraints on the
properties of dark matter halos. A detailed discussion of the results
can be found in Hoekstra et al. (2004). The analysis presented here
uses only $R_C$ imaging data from the RCS, and therefore lacks
redshift information for the individual lenses. Nevertheless, these
measurements allow us to place tight constraints on the extent and
masses of dark matter halos.

In our maximum likelihood analysis we consider $r_s$ and $V_{200}$ (or
equivalently the mass $M_{200}$) as free
parameters. Figure~\ref{size_nfw} shows the joint constraints on
$V_{200}$ and scale radius $r_s$ for a fiducial galaxy with $L_{\rm
B}=10^{10}h^{-2}L_{{\rm B}\odot}$, when we use an NFW profile for the
galaxy dark matter profile. Numerical simulations of CDM, however,
show that the parameters in the NFW model are correlated, albeit with
some scatter. Hence, the simulations make a definite prediction for
the value of $V_{200}$ as a function of $r_s$. The dotted line in
Figure~\ref{size_nfw} indicates this prediction. If the simulations
provide a good description of dark matter halos, the dotted line
should intersect our confidence region, which it does.

This result provides important support for the CDM paradigm, as it
predicts the correct ``size'' of dark matter halos. It is important to
note that this analysis is a direct test of CDM (albeit not
conclusive), because the weak lensing results are inferred from the
gravitational potential at large distances from the galaxy center,
where dark matter dominates.  Most other attempts to test CDM are
confined to the inner regions, where baryons are, or might be,
important.

Another prediction from CDM simulations is that halos are not
spherical but triaxial instead. We note, however, it is not completely
clear how the interplay with baryons might change this. For instance,
Kazantzidis et al. (2004) find that similations with gas cooling are
significantly rounder than halos formed in adiabatic simulations, an
effect that is found to persist almost out to the virial radius.
Hence, a measurement of the average shape of dark matter halos is
important, because both observational and theoretical constraints are
limited. Weak gravitational lensing is potentially one of the most
powerfuls way to derive constraints on the shapes of dark matter
halos. The amount of data required for such a measurement, however, is
very large: the galaxy lensing signal itself is tiny, and now one
needs to measure an even smaller azimuthal variation. We also have to
make assumptions about the alignment between the galaxy and the
surrounding halo. An imperfect alignment between light and halo will
reduce the amplitude of the azimuthal variation detectable in the weak
lensing analysis.  Hence, weak lensing formally provides a lower limit
to the average halo ellipticity.

Hoekstra et al. (2004) attempted such a measurement, again using a
maximum likelihood model. They adopted a simple approach, and assumed
that the (projected) ellipticity of the dark matter halo is
proportional to the shape of the galaxy: $e_{\rm halo}=f e_{\rm
lens}$. This yielded a a best fit value of $f=0.77^{+0.18}_{-0.21}$
(68\% confidence), suggesting that, on average, the dark matter
distribution is rounder than the light distribution. Note, however,
that even with a data set such as the RCS, the detection is marginal.
A similar, quick analysis of imaging data from the CFHTLS and
VIRMOS-Descart surveys give lower values for $f$, suggesting that the
RCS result is on the high side. 

Recently, an independent weak lensing measurement of halo shapes was
reported by Mandelbaum et al. (2005), based on SDSS observations.  For
the full sample of lenses they do not detect an azimuthal variation of
the signal, which is somewhat at odds with the Hoekstra et al. (2004)
findings.  However, as Mandelbaum et al. (2005) argue, the comparison
is difficult at best, because of different sensitivity to lens
populations, etc. and differences in the analyses. However, the
approach used by Mandelbaum et al. (2005) has the nice feature that it
is more `direct', compared to the maximum likelihood approach.  The
latter `always gives an answer', but in our case it is difficult to
determine what scales or galaxies contribute most to the signal.
Interestingly, Mandelbaum et al. (2005) also split the sample into
blue (spiral) and red (elliptical) galaxies. The results suggest a
positive alignment between the dark matter halo and the brightest
sample of ellipticals, whereas the spiral galaxies might be aligned
perpendicular to the disks. Although the signal in both cases is
consistent with 0, it nevertheless provides an interesting that
deserves further study.

\section{Outlook}

The results presented in the previous two section provide a crude
picture of what weak lensing studies of galaxy halos can accomplish
with current data sets. For galaxy-galaxy lensing studies both the RCS
and SDSS data sets provide the most accurate results, with SDSS having
the advantage of a larger number of galaxies with (photometric)
redshift information. Even though these are early results (the
galaxy-galaxy lensing was first detected less than a decade ago),
already we can place interesting constraints on the properties of dark
matter halos and the stellar contents of galaxies.

Much larger surveys have started. For instance, the second generation
RCS aims to image almost 850 deg$^2$ in $g',r',z'$. These data provide
more than an order of magnitude improvement over the results discussed
in these proceedings. The KIlo Degree Survey (KIDS) will start
observations soon using the VLT Survey Telescope. This survey will
image $\sim 1500$ deg$^2$ (to a depth similar to that of RCS2) in five
filters, thus adding photometric redshift information for most of the
lenses. The Canada-France-Hawaii-Telescope Legacy Survey will also
provide important measurement of the lensing signal induced by
galaxies. It is much deeper than RCS2 or KIDS, but will survey a
smaller area of $\sim 170$ deg$^2$, with cosmic shear measurements as
the primary science driver. Nevertheless its signal-to-noise ratio of
the measurements will be comparable to the RCS2, but it will have the
advantage of accurate photometric redshift information from the 5
color photometry.  Thanks to its added depth, it is also well suited
to study the evolution of galaxy properties.  Dedicated survey
telescopes such as PanSTARRS or the LSST will image large portions of
the sky, thus increasing survey area by another order of magnitude to
a significant fraction of the sky.

One of the most interesting results from these projects will be a
definite measurement of the average shape of dark matter halos. We can
expect much progress on this front in the next few years. Although
there is much reason for optimism, we also need to be somewhat
cautious: the accuracy of the measurements is increasing rapidly, but
it is not clear to what extent the interpretation of the results can
keep up.  The early results, presented here, have statistical errors
that are larger than the typical model uncertainty. However, as
measurement errors become are significantly smaller, it becomes much
more difficult to interpret the measurements: some more subtle effects
arising from neigbouring galaxies or satellite galaxies can no longer
be ignored. Instead, it will become necessary to compare the lensing
measurements (i.e., the galaxy-mass cross-correlation function as a
function of galaxy properties) to results of simulations directly.
These future studies will provide unique constraints on models of
galaxy formation as they provide measures of the role dark matter
plays in galaxy formation.

\section*{Acknowledgements} Much of the work presented here would
not have been possible without the efforts of the members of the RCS
team. In particular I acknowledge the work of Paul Hsieh and Hun Lin
on the photometric redshifts and Howard Yee and Mike Gladders on their
work on RCS in general.



\end{document}